\documentclass{article}
\usepackage{arxiv,epsfig}
\usepackage{amsmath}
\usepackage{amssymb}
\usepackage{amsthm}
\usepackage{textcomp}
\usepackage{graphicx}%
\usepackage{hyperref}
\hypersetup{
    bookmarks=true,         %
    unicode=false,          %
    pdftoolbar=true,        %
    pdfmenubar=true,        %
    pdffitwindow=false,     %
    pdfstartview={FitH},    %
    pdftitle={},    %
    pdfauthor={},     %
    pdfsubject={},   %
    pdfcreator={},   %
    pdfproducer={}, %
    pdfkeywords={}{}{}{}{}, %
    pdfnewwindow=true,      %
    colorlinks=true,       %
    linkcolor={blue},          %
    citecolor=blue,        %
    filecolor=blue,      %
    urlcolor=blue           %
}
\usepackage{tikz}
\usetikzlibrary{arrows}
\definecolor{mycolor}{rgb}{0.1, 0.5, 0.2}
\DeclareMathOperator*{\argmax}{arg\,max}

\newcommand{\Real}{{\mathbb R}}
\DeclareMathOperator{\Tr}{Tr}

\title{Spatial Noise-aware Temperature Retrieval from Infrared Sounder Data}

\author{
  David Malmgren-Hansen \\
  Department of Applied Mathematics and Computer Science \\
  Technical University of Denmark \\
  Denmark\\
  \And
  Valero Laparra \\
  Image Processing Laboratory \\
  Universitat de Val{\`e}ncia\\
  Val{\`e}ncia, Spain\\
  \texttt{valero.laparra@uv.es} \\
  \And
  Allan Aasbjerg Nielsen \\
  Department of Applied Mathematics and Computer Science \\
  Technical University of Denmark \\
  Denmark\\
  \And
  Gustau Camps-Valls \\
  Image Processing Laboratory \\
  Universitat de Val{\`e}ncia\\
  Val{\`e}ncia, Spain\\
  \texttt{gcamps@uv.es} \\
}

\begin{document}

\begin{center}
    ©IEEE. ACCEPTED FOR PUBLICATION IN IEEE IGARSS 2017. DOI 10.1109/IGARSS.2017.8126882\footnote{
©IEEE. Personal use of this material is permitted.  Permission from IEEE must be obtained for all other users,including reprinting/republishing this material for advertising or promotional purposes, creating new collective works for resale or redistribution to servers or lists, or reuse of any copyrighted components of this work in other works.  DOI: 10.1109/IGARSS.2017.8126882.
}
\end{center}

\maketitle

\begin{abstract}
In this paper we present a combined strategy for the retrieval of atmospheric profiles from infrared sounders. The approach considers the spatial information and a noise-dependent dimensionality reduction approach. The extracted features are fed into a canonical linear regression. We compare Principal Component Analysis (PCA) and Minimum Noise Fraction (MNF) for dimensionality reduction, and study the compactness and information content of the extracted features. Assessment of the results is done on a big dataset covering many spatial and temporal situations. %
PCA is widely used for these purposes but our analysis shows that one can gain significant improvements of the error rates when using MNF instead. In our analysis we also investigate the relationship between error rate improvements when including more spectral and spatial components in the regression model, aiming to uncover the trade-off between model complexity and error rates. %
\end{abstract}
\section{Introduction}
\label{sec:intro}
\begin{flushright}
{\em ``Perfection is achieved not when there is nothing more to add, but when there is nothing more to take away.''\\
--- Antoine de Saint-Exupéry: Terre des hommes.}
\end{flushright}
Temperature and water vapour atmospheric profiles are essential meteorological parameters for weather forecasting and atmospheric chemistry studies. Observations from high spectral resolution infrared sounding instruments on board of satellites provide for retrieval of such profiles. However, it is not trivial to retrieve the full information content from radiation measurements; accordingly, improved retrieval algorithms are desirable to achieve optimal performance for existing and future infrared sounding instrumentation. 

EUMETSAT, NOAAA, NASA and other agencies are continuously developing product processing facilities to obtain L2 atmospheric profile products from infrared hyperspectral radiance instruments, such as IASI. One of the retrieval techniques commonly used in L2 processing is based on linear regression, which is a valuable and very computationally efficient method.
It consists of performing a canonical least squares linear regression on top of the data projected onto the first principal components or Empirical Orthogonal Functions (EOF) --known in statistics as PCA-- of the measured brightness temperature spectra (or radiances) and the atmospheric state parameters. To further improve the results of this scheme for retrieval, nonlinear statistical retrieval methods, as well as nonlinear pre-processing methods \cite{Arenas-Garcia201316}, can be applied as an efficient alternative to more costly optimal estimation (OE) schemes. These methods have proven to be valid in retrieval of temperature, dew point temperature (humidity), and ozone atmospheric profiles when the original data are used~\cite{CAMPS2012,CampsValls10eumigarss}. However, they are costly to train and do not consider spatial correlation between radiances neither the noise information.

Recently, in \cite{Garcia17compress}, a high improvement on the performance of retrieval methods was reported when applying standard compression algorithms to the images. Although this result may appear counter-intuitive since compression implies reduction on the amount of information in the images, a certain level of compression is actually useful because: 1) compression removes information but also noise, and it could be useful to remove the components with low signal-to-noise ratio.; and 2) spatial compression introduces in a simple way information about the neighboring pixels. %
The use of Minimum Noise Fractions (MNF) employed here is a simpler and more mathematically elegant way to take advantage of both properties simultaneously. MNF is specifically designed to sort components according to the signal-to-noise ratio (SNR) \cite{green1988transformation}. The way we apply MNF here also enforces the inclusion of spatial information as noise is estimated by the residuals of fitting a quadratic surface locally. %
In this work we compare the effect of using PCA or MNF when retrieving temperature profiles using IASI data. We will show that MNF is better suited for this task. Moreover since PCA and MNF are both linear and unsupervised transformations, using MNF do not introduce any modification in the data processing pipeline. %

The remainder of the work is organized as follows. Section \S2 describes the data set collected and the pre-processing for dimensionality reduction and spatial filtering. Section \S3 reviews the two decomposition methods used in the work. Section \S4 gives empirical evidence of performance of the proposed scheme for spatial, noise-aware retrieval of atmospheric parameters. We conclude in \S5 with some remarks and outline for the further work.

\section{Data description}
\label{sec:data}

The Infrared Atmospheric Sounding Interferometer (IASI) data are point measurements of approximately $25$ $km$ diameter with $8461$ spectral components, ranging in the infrared emission spectra from $645$ to $2760$ $cm^{-1}$ with $0.25$ $cm^{-1}$ resolution. The dataset collected for this paper consists of 4 consecutive orbits from august 2013 of which the first three are used for training the regression model and the last is used for testing. 

In our problem we follow the same scheme proposed in \cite{Garcia17compress}. First we remove certain bands from the spectrum that do not contains useful information for retrieval reducing the data to $4699$ spectral components. Although the longitudinal distance between acquisition points increases towards equator we can reshape each orbit into a rectangular grid of $1530 \times 60$ elements. By doing so, data can be treated as an image, taking advantage of spatial relations. The dimensionality reduction transformations are calculated on the training set and applied to both the training and testing datasets. 

\section{Decomposition Methods}
\label{sec:decomp}

In our analysis we consider two orthogonal transformations, PCA \cite{hotelling1933analysis} and MNF \cite{green1988transformation}. Notationally, given an observation data matrix $\mathbf{X}\in\Real^{n\times d}$ with $n$ pixels of $d$ dimensions, we aim to find a transformation to a lower dimensional representation, $d'<d$, such that the projected data preserves most of the `information' of the input. Solutions offered by both PCA and MNF are found by solving an eigenvalue problem but where the PCA finds a solution with eigenvectors in the columns of $\mathbf{W}\in\Real^{d\times d'}$ in direction of maximum variance, the MNF looks for the eigenvectors that minimize the noise fraction, or equivalently maximizes the signal-to-noise ratio \cite{nielsen2011kernel,gomez2011explicit}:
\begin{equation}\label{eq:raycoeffs}
\begin{split}
    \text{PCA}: \mathbf{W_*} & = \argmax_{\mathbf{W}} \left \{ \Tr \left (  {\frac{\mathbf{W}^\top\mathbf{X}^\top\mathbf{X}\mathbf{W}}{\mathbf{W}^\top\mathbf{W}}}  \right ) \right \} \\ \\
    \text{MNF}: \mathbf{W_*} & = \argmax_{\mathbf{W}} \left \{ \Tr \left ( \frac{\mathbf{W}^\top\mathbf{X}^\top\mathbf{X}\mathbf{W}}{\mathbf{W}^\top\mathbf{X}_N^\top\mathbf{X}_N\mathbf{W}} \right ) \right \},
\end{split}
\end{equation}
where $\mathbf{X}$ is our data matrix with each row representing a sample of a infrared spectrum and with columns corresponding to the number of spectral components. $\mathbf{X}_N$ is the corresponding noise estimation of each sample in $\mathbf{X}$. The resulting set of vectors from the PCA decomposition are orthogonal as opposed to the MNF solution which obtains orthogonality with respect to the noise covariance.

\if false
defined as: %

\begin{equation}\label{eq:mnf}
\begin{split}
    S & = S_S + S_N \\ \\
    NF & = \frac{w^TS_Nw}{w^TSw} 
\end{split}
\end{equation}

\begin{equation}\label{eq:raycoeffs}
\begin{split}
    PCA: R(w) & = \frac{w^TX^TXw}{w^Tw}  \\ \\
    MNF: R(w) & = \frac{1}{NF}  = \frac{w^TS_Nw}{w^TSw} = \frac{w^TX^TXw}{w^TX_N^TX_Nw} 
\end{split}
\end{equation}

\fi

If the noise covariance matrix is known, it can be used in the MNF estimation. Often it is not the case and it has to be estimated from data. Common ways to do noise estimation in image analysis include local mean subtraction, or taking the residuals from a plane or paraboloid fit on every pixel position in the image. We follow the latter approach for our analysis with a $3\times 3$ paraboloid residual kernel implemented as a filtering operation~\cite{nielsen1999extension}. %

\begin{figure}[htb]
\centering
\includegraphics[width=0.8\linewidth]{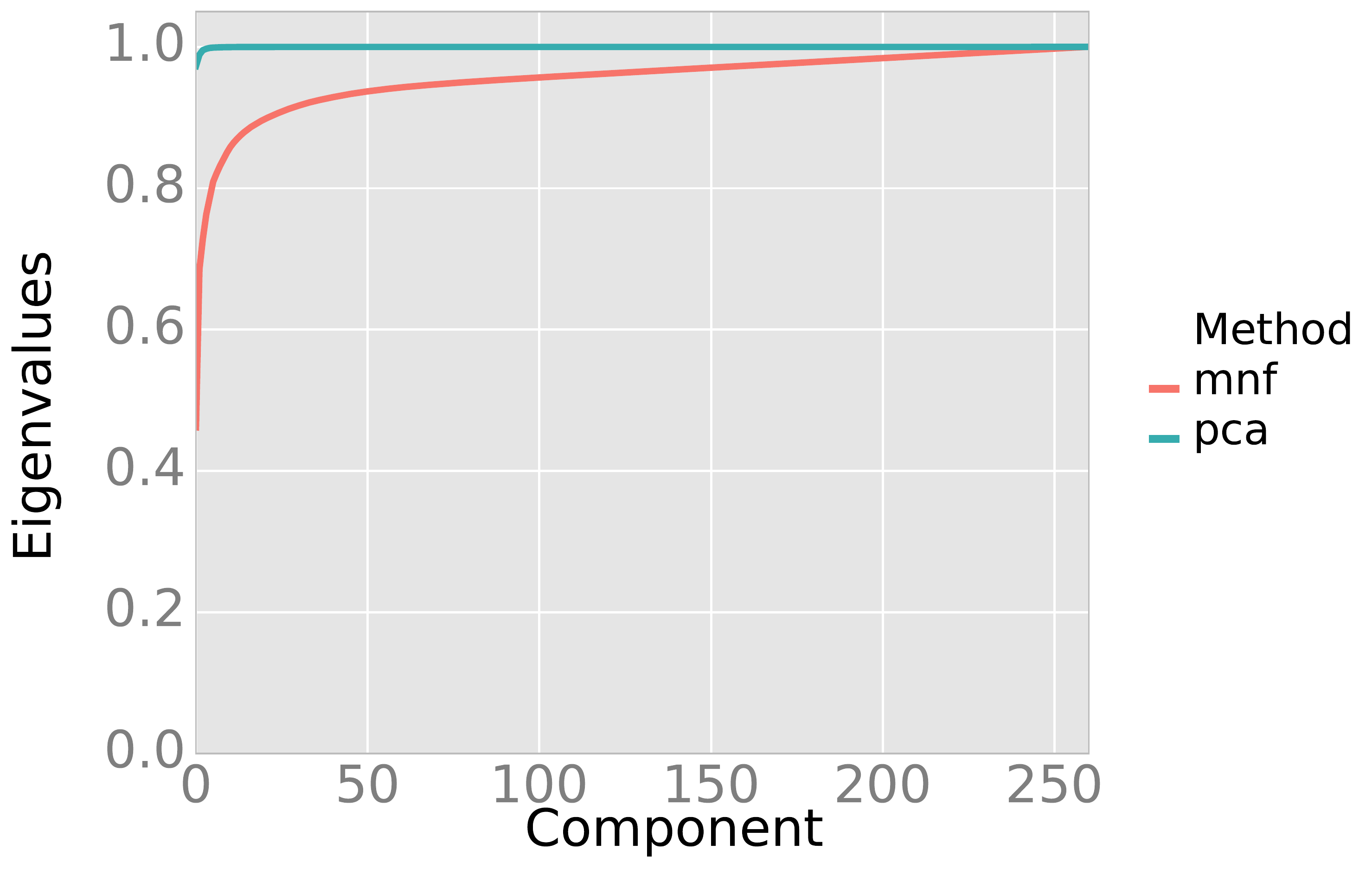}
\caption{Cummulated normalized eigenvalues.}
\label{fig:eig}
\end{figure}

\begin{figure}[htb]
\centering
\includegraphics[width=0.8\linewidth]{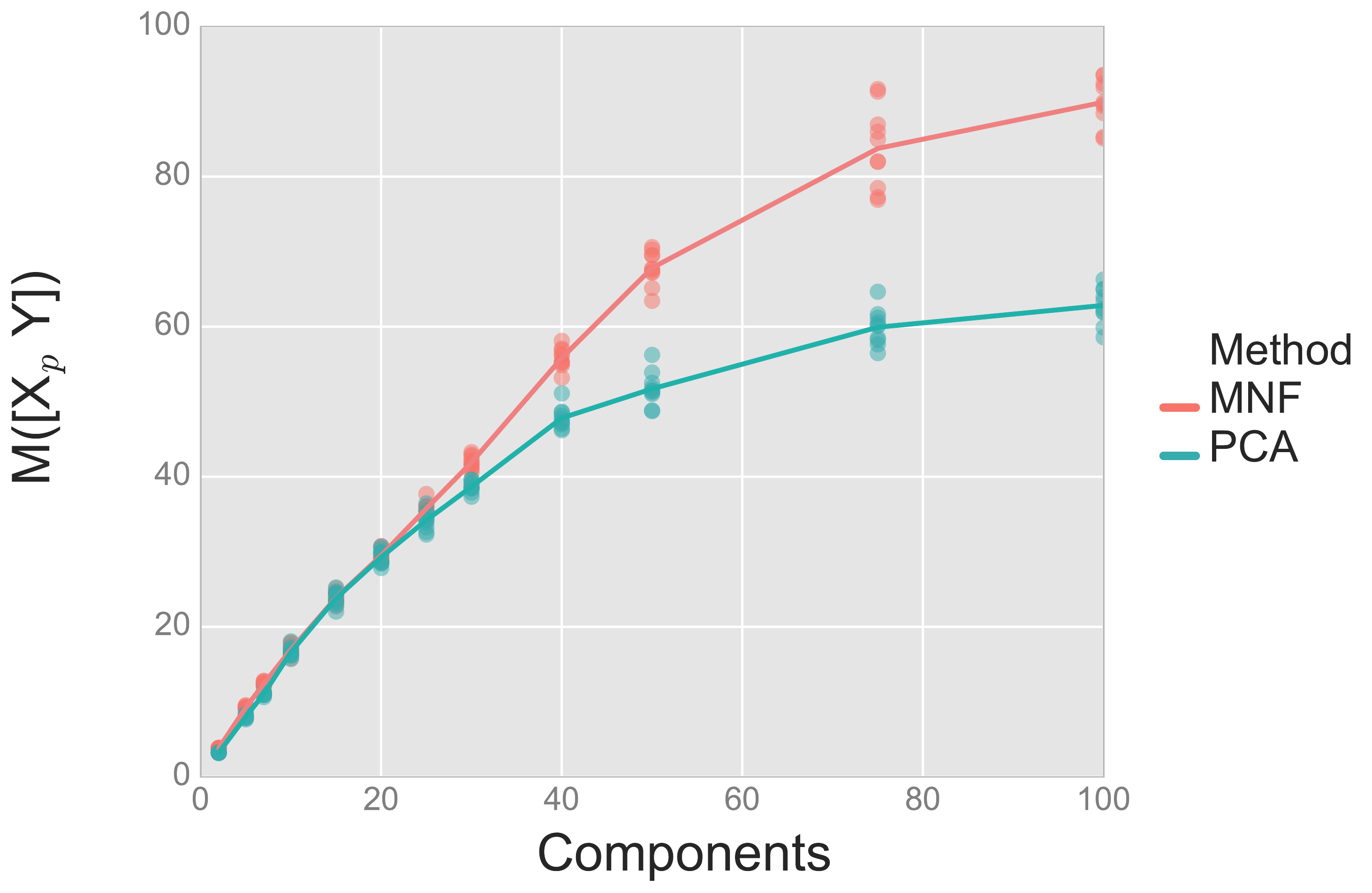}
\caption{Multi-information between the input and output components for PCA and MNF. Ten realizations have been made for each method while including different amount of spectral components. Lines denote the mean trend of the results.}
\label{fig:MI}
\end{figure}

The cumulative and normalized eigenvalues for both methods are shown in Fig.~\ref{fig:eig}. For PCA they represent the percentage of explained total variance and it is seen that 99\% explained variance is obtained within the first $5$ components. For the MNF, eigenvalues represent the signal fraction for each component \cite{green1988transformation} and less than 80\% signal fraction is obtained from the first 5 components. Although this could be seen as an disadvantage one has to take into account that PCA might keep the noise information too. Therefore how the eigenvalues relates to the information necessary to predict the temperature profiles $\mathbf{Y}\in\Real^o$ is less straight forward to estimate. We analyze in Fig.~\ref{fig:MI} this relation by using the concept of {\em multiinformation}~\cite{Studeny1998} (also known as {\em total correlation}). We show the amount of multiinformation, i.e. shared information, between the projected inputs $\mathbf{X}_p = \mathbf{XW}$ and the outputs $\mathbf{Y}$ using different amount of input components for each decomposition method. These values have been computed using RBIG method (\cite{RBIG11}). We have followed similar procedure as in \cite{laparrcr2015spatial} where the amount of information contained by spatial and spectral components was analyzed for several sensor configurations. In this case, we are analyzing only the spectral information, yet including the variable to predict $\mathbf{Y}$ (temperature profiles). Fig.~\ref{fig:MI} shows the multiinformation results for PCA and MNF. Although it also includes the redundant information of the inputs, this measure can be seen as an approximation of the information of the output that we can be obtained from the input. Note that even that MNF is not specifically designed to maximize this information, the multiinformation is bigger for MNF when using the same number of input components than for PCA. We will see in the experiments section how this behavior gives raise to improved retrieval performance.

\begin{figure*}
\centering
\includegraphics[width=\textwidth]{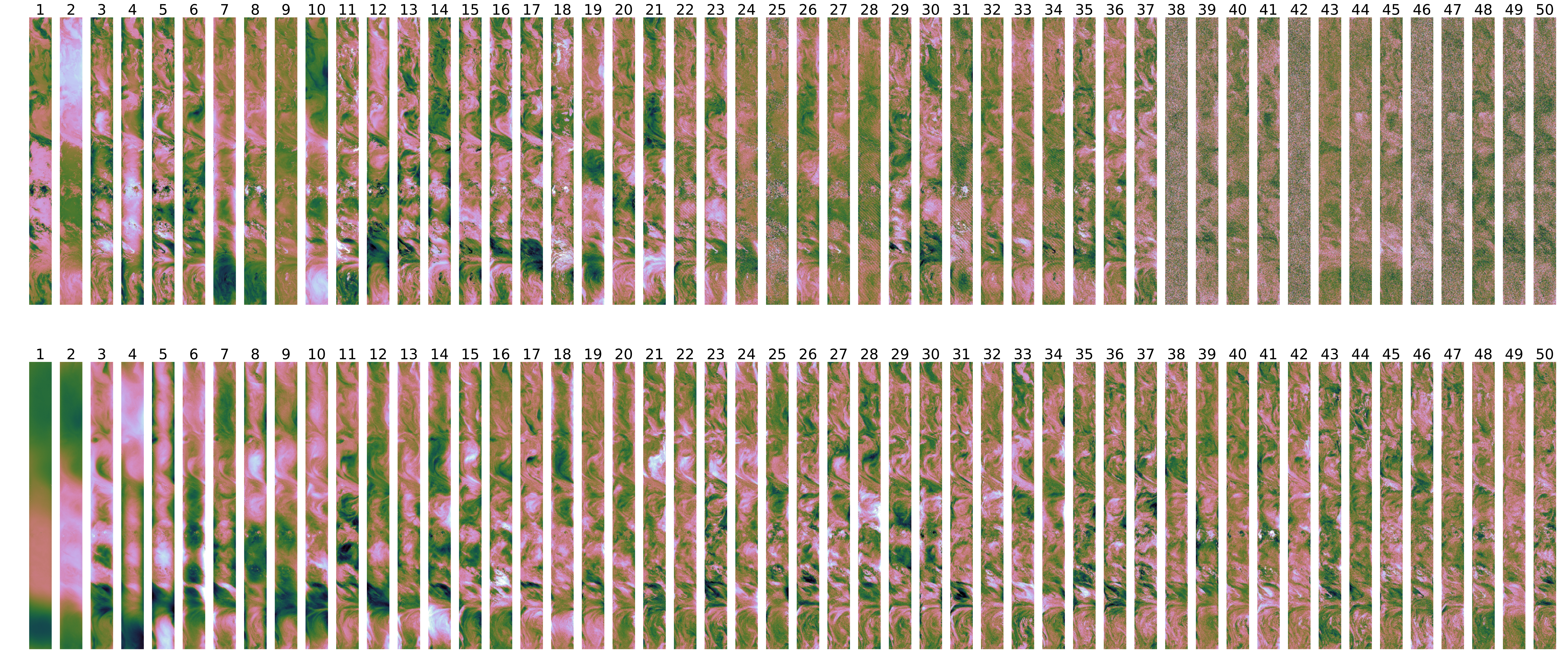}
\caption{Each strip is half an orbit projected onto the different components (PCA top, MNF bottom).}\label{fig:dataproj}
\end{figure*}

As suggested in \cite{Garcia17compress}, to improve the retrieval performance it is important to remove noise from the data and to include spatial information. Fig.\ \ref{fig:dataproj} illustrates the ability of MNF to do so. We show half orbit of data from the test set projected onto each of the 50 first components from the PCA (top row) and MNF (bottom row) decomposition. It is clear from this figure that MNF obtains smoother are less noisy projections than PCA. For instance component $38$ from the PCA projection seem to contain less structure than the three following projections. This indicates that some noise components in the data have higher variance than other signal components. This behaviour repeats above the first $50$ PCA components, whereas the MNF projections represents spatially smooth information in early components and gradually increase to finer details for higher components.

\section{Experimental Results}
\label{sec:exp}

The goal of our experiments is two-fold, first to compare the effect of using PCA or MNF in the retrievals, and furthermore to uncover the trade-off between prediction performance and the number of spectral components included for each method. 
Dimensionality reduction is important to limit the computational load but choosing the appropriate number of components to keep is less straightforward. A lower computational load can be traded for larger amounts of training data so overfitting is prevented. Alternatively the lower number of data dimensions can enable the use of computationally heavier non-linear models such as Kernel Ridge Regression, which has been shown to improve performance for retrieval in infrared sounder data \cite{Garcia17compress}. 

As well as the influence of spectral sampling in temperature profile modelling we include experiments for different sizes of pixel neighborhood sampling as studies suggest this can be beneficial \cite{laparrcr2015spatial}. This means that we model the temperature profile of one sample in the IASI data from the sample plus a neighborhood of samples around it. For quadratic neighborhoods the increase of size will also lead to quadratic increase in computational load and it is therefore relevant to limit it.

\begin{figure}
\centering
\includegraphics[width=\linewidth]{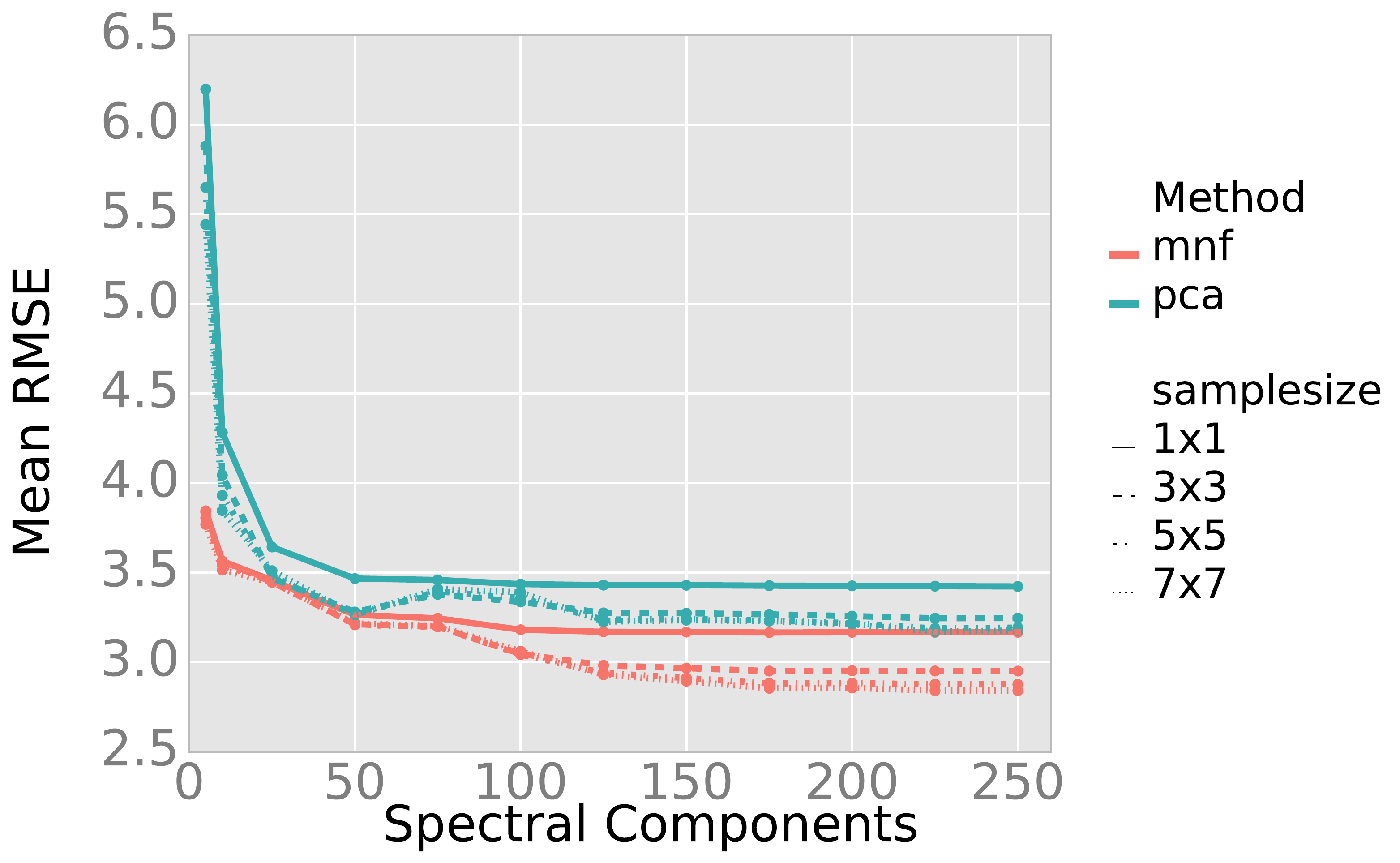}
\caption{Mean of RMSE on temperature profiles for different decomposition methods and spatial sample sizes, when including an increasing amount of spectral components. The Y-axis is the mean of the root-mean-square-error on prediction with $o=90$ i.e. predition of 90 altitudes in the atmosphere given by their pressure level.}\label{fig:resultplot}
\end{figure} 

\begin{figure}[htb]
\centering
\includegraphics[width=\linewidth]{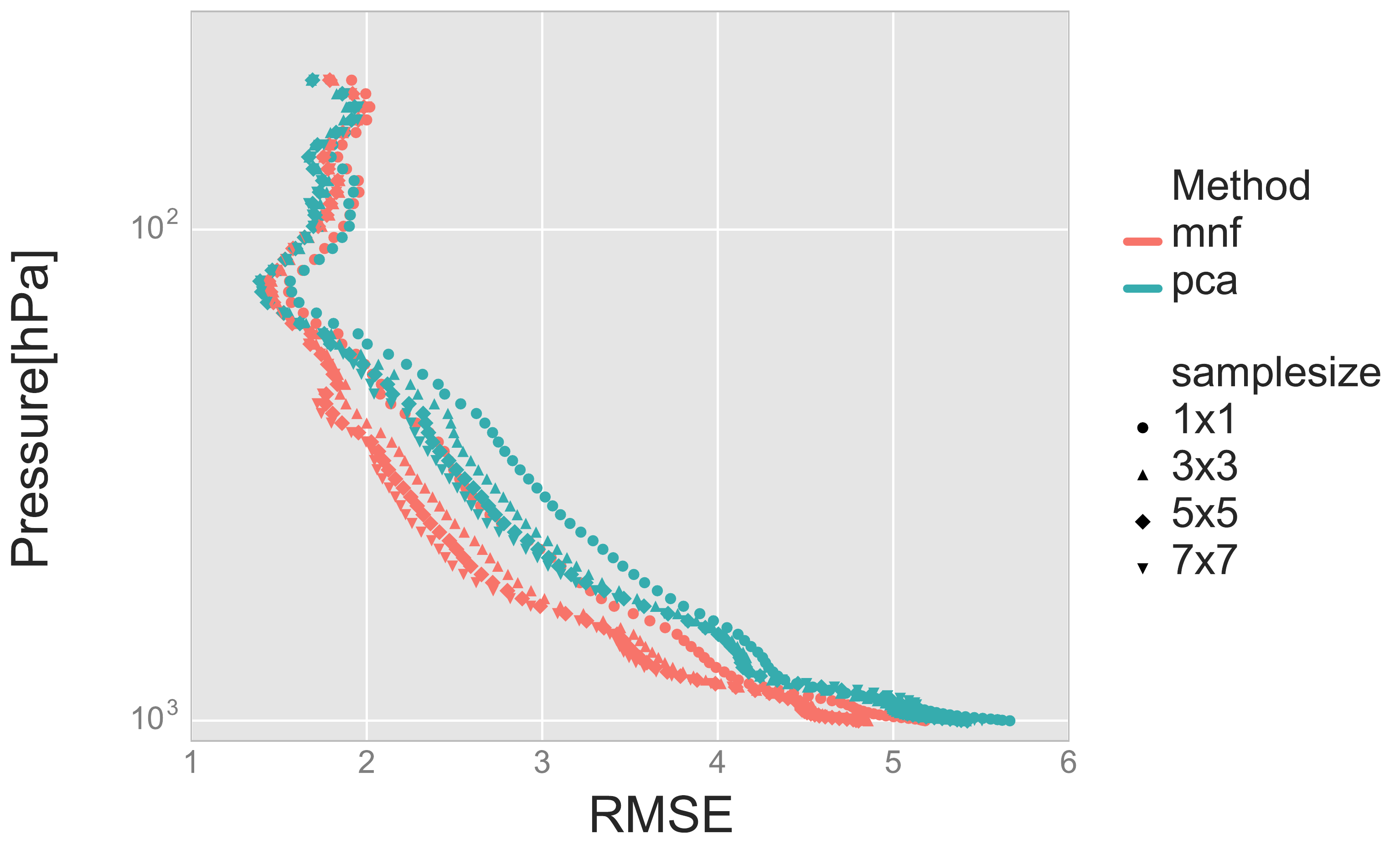}
\caption{RMSE on temperature profiles for 90 different altitudes shown by the air pressure (y-axis). In this experiment 175 spectral components was used to fit the Linear Regression model for 3 different neighborhood sampling size on PCA and MNF transformation. Measurements on clouds yields typically higher error rates for lower altitudes. Note that in this experiment nothing has been done to filter away measurements dominated by clouds.}
\label{fig:resultprof}
\end{figure}

In Fig.\ \ref{fig:resultplot} the results from our experiments are shown. It is seen that the RMSE improvements converges after approximately 125 spectral components. The results also show that there is a significant improvement including neighborhood pixels in the modelling of temperature profiles, but that the improvement decreases going towards larger neighborhood sizes. Figure~\ref{fig:resultprof} shows the resulting RMSE over the temperature profile for using 175 spectral components in the Linear regression model. Our analysis suggests to use between $125-175$ spectral components from a MNF decomposition and a pixel neighborhood sampling size of $3\times 3$ or $5\times 5$ when performing Linear Regression on this type of data.

\section{Conclusions}
\label{sec:con}

This paper showed that using MNF is a simple and mathematically elegant way of removing the noise in the signal and at the same time taking into account spatial information. These two properties have been suggested previously as an important point when dealing with this particular data \cite{Garcia17compress}. Both effects can be observed in Fig.~\ref{fig:dataproj}, the selected features by the MNF are less noisy and spatially softer than the ones found by PCA. We want to stress the fact that substituting PCA by MNF would not change the processing pipeline. PCA and MNF are both linear transformations so only the values of the projecting vectors should be changed. Moreover, unlike other solutions as PLS \cite{Wold1966}, PCA and MNF are unsupervised methods, i.e.\ are not fitted for predicting an specific variable. Therefore, although we here show the results for a particular variable (i.e. temperature), it is expected that the improvement would be consistent for the retrieval of other variables.   

\section*{Acknowledgements}
The research was funded by the European Research Council (ERC) under the ERC-CoG-2014 SEDAL project (grant agreement 647423), and the Spanish Ministry of Economy and Competitiveness (MINECO) through the projects TIN2015-64210-R and TEC2016-77741-R.

\small
\bibliographystyle{IEEEbib}
\bibliography{references}

\end{document}